\documentclass{WileyMSP-template}
\usepackage{amsmath,mathtools}
\usepackage{textcomp}
\usepackage{braket}
\usepackage{amssymb,bm}
\usepackage{hyperref}% add hypertext capabilities
\begin{document}

\pagestyle{fancy}
\rhead{\includegraphics[width=2.5cm]{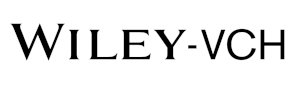}}

\title{Moiré folded helical states at the interfaces of heterostructures}

\maketitle

% Author: Please give full first and last names for authors and include * after the name of all corresponding authors

\author{Paula Mellado*}

% Dedication

\dedication{}

% Affiliations: Please provide adacemic titles (Prof. or Dr.) for all authors where applicable, and include an institutional email address for all corresponding authors
\begin{affiliations}
Professor Paula Mellado\\
Facultad de Ingeniería y Ciencias,\\ Universidad Adolfo Ibáñez, Santiago, Chile.\\
paula.mellado@uai.cl

\end{affiliations}

% Keywords: Please provide a minimum of three and a maximum of seven keywords, separated by commas


% Abstract should be written in the present tense and impersonal style (i.e., avoid we), and be at most 200 words long
\begin{abstract}

A minimal model of a graphene–topological-insulator heterostructure is considered, where a moiré superlattice modulates the Rashba spin–orbit interaction (SOC). In the spin-degenerate, spin–orbit-free limit, the reduced Brillouin zone contains flat, spin-degenerate moiré minibands, with periodicity determined by superlattice folding. The inclusion of SOC lifts the spin degeneracy and reduces the effective spectral periodicity by a factor of two. Through SOC, the moiré potential entangles spin, sublattice, and leg degrees of freedom, reshaping the miniband structure in momentum space and generating emergent helicity spectral functions. As the Rashba coupling is renormalized by the moiré pattern, it induces helicity fragmentation, in which the helicity weight is distributed across a dense manifold of moiré minibands, forming an extended network of helicity-carrying states and significantly enhancing helicity fluctuations at the bare-response level. The emergence of Dirac-like miniband crossings at finite SOC demonstrates that moiré heterostructures can support relativistic quasiparticles through band reconstruction. This model provides a microscopic mechanism by which proximity-induced spin–orbit coupling can be amplified via moiré engineering.

\end{abstract}

% Text: Please use section headings and subheadings as specified below. For communications, all section headings apart from Experimental Section should be removed
% Please make the first reference to a display item bold: \textbf{Figure 1}
% Do not abbreviate Figure, Equation, etc.; display items are always singular, i.e., Figure 1 and 2.
% Equations are always singular, i.e., Equation 1 and 2, and should be inserted using the {equation} environment, not as graphics
% Please do not use footnotes in the text, additional information can be added to the Reference list.

\section{\label{sec:intro}Introduction}
Materials with intrinsically weak spin–orbit coupling (SOC) do not exhibit spin–momentum locking or spin-driven collective states \cite{manchon2015new,soumyanarayanan2016emergent}. An essential approach to introduce SOC into high-mobility systems is to exploit proximity effects in van der Waals heterostructures, which allows a low-SOC material to inherit SOC from an adjacent strong-SOC layer \cite{avsar2020colloquium,sierra2021van}. In van der Waals heterostructures, interfacial spin–orbit coupling  originates from broken inversion symmetry combined with strong atomic SOC in one of the constituent layers. In this regard, graphene/three-dimensional topological insulator (TI) heterostructures are promising, as they combine Dirac quasiparticles, strong spin–momentum locking, adjustable electrostatics, and atomically clean interfaces \cite{hasan2010colloquium}. Experimental and theoretical investigations of graphene/TI heterostructures have shown that TI surface states can induce strong spin–orbit coupling in graphene, leading to spin splitting, anisotropic spin textures, and gate-tunable helical transport behavior \cite{zollner2021heterostructures,lee2015proximity,zhang2014proximity,song2018spin,kiemle2022gate,khokhriakov2020gate}. Tight-binding and ab-initio calculations reveal that the induced SOC is highly sensitive to interface symmetry, stacking configuration, and sublattice alignment, typically producing a complex combination of Rashba, valley–Zeeman, and pseudospin-dependent SOC terms \cite{gmitra2018proximity,gani2020proximity,zollner2016theory}. Despite these advances, a central unresolved question is whether the observed SOC simply reflects a passive inheritance of the TI’s spin texture, or whether it can instead be actively enhanced and reprogrammed via additional interfacial structural modulations \cite{li2019twist,premasiri2019tuning,wakamura2019spin}.

A promising approach to addressing this issue considers the formation of a moiré superlattice at the graphene/TI interface. Such moiré patterns can be induced by lattice mismatch, relative twist, or commensurate stacking over large unit cells \cite{yankowitz2019tuning}. It has been shown in several twisted heterostructures that moiré patterns produce flattened electronic minibands and enhance interband hybridization, thereby amplifying correlation effects \cite{bistritzer2011moire}. In several graphene moiré platforms, including twisted bilayer graphene \cite{cao2018correlated} and graphene on transition-metal dichalcogenides \cite{andrei2021marvels}, such effects are known to substantially modify electronic, magnetic, and topological characteristics \cite{xiao2020moire}. In graphene/TI heterostructures, where proximity already endows graphene with spin–orbit coupling \cite{zhang2014proximity}, the moiré pattern produces spatially varying interlayer and intersublattice hybridization that can couple strongly to spin–orbit effects \cite{tseng2022anomalous,zollner2021bilayer}.

In this article, we introduce a minimal model that isolates the combined effects of proximity-induced SOC and moiré modulated interband hopping in graphene based heterostructures. Rather than focusing on material-specific details, we employ a toy model approach inspired by recent works on moiré ladders and incommensurate heterostructures \cite{mellado2025sliding,gao2022symmetry,mellado2025excitonic,munoz2026emergence}. We investigate how moiré modulation influences a spin–orbit coupled heterostructure, focusing on whether the moiré potential can enhance spin–momentum locking and promote the emergence of a collective instability in the system.  To that end, we characterize the dominant fluctuation channels of this system, focusing on helicity rather than conventional charge or spin order. We analyze the bare susceptibilities associated with helicity density and helicity current operators. We show that while helicity spectral functions vanish in the absence of SOC, the  static susceptibilities become strongly enhanced once SOC and moiré hybridization coexist.  Our results open a new route toward designing helical and spin orbit driven phases in van der Waals heterostructures through controlled structural modulation.

\section{\label{sec:model}Model}
We model graphene/TI heterostructures as a ladder system, where the top and bottom legs represent the top and bottom surfaces of the heterostructure, as illustrated in Fig.\ref{f1}(a). Each ladder has two sites (A and B representing two different atoms) per unit cell, and the composite ladder unit cell contains a total of four sites $\{A_1,B_1,A_2,B_2\}$ (subscripts $1,2$ label legs).  The composite cell has a length equal to $2a$, and the original reciprocal period is \(
G_0=\pi/a\). Henceforth, we choose half of the legs lattice constant as the length unit. 

The two-leg ladder is subject to  moiré modulation due to the slight dimerization in leg-2 as shown in Fig.\ref{f1}(a).  The two legs share the same microscopic lattice periodicity, such that no geometric moiré pattern emerges directly at the level of the underlying atomic lattice. Instead, the moiré superstructure is incorporated phenomenologically via a long-wavelength modulation of the inter-leg (rung) tunneling amplitude. This modulation can be interpreted as an emergent consequence of spontaneously broken translational symmetry, which may arise, for example, from a soft phonon mode that induces relative phase shifts or displacements between neighboring chains.  A microscopic realization of this mechanism has been proposed for $\rm HfTe_3$ \cite{mellado2025sliding,mellado2025excitonic}, where first-principles calculations reveal a chain-dependent structural instability that gives rise to such a modulation. 

The microscopic tight-binding Hamiltonian for spinful fermions formulated in the extended Brillouin zone (EBZ), contains eight bands due to leg, orbital, and spin degrees of freedom. The fermionic operators are defined as
\(
c_{j,\ell,\sigma}
\),
where $\ell=1,2$ labels the leg and $\sigma=\uparrow,\downarrow$ is the spin quantized along the $z$ axis. The Hamiltonian is written as
\begin{equation}
H = H_K + H_{\mathrm{SOC}},
\label{eq:allH}
\end{equation}
with $H_K=H_{\parallel} + H_{\perp}$ the contribution from the kinetic energy along the ladder. It comprises an intra-leg nearest neighbor hopping term, $H_{\parallel}$ given by
\begin{equation}
H_{\parallel}
= \sum_{j,\ell,\sigma}
-t_\parallel\left(
c^\dagger_{j+1,\ell,\sigma} c_{j,\ell,\sigma} + \mathrm{h.c.}
\right),
\label{eq:hpar}
\end{equation}

In Eq.\ref{eq:hpar}, the tunneling amplitude in leg-1 $t_\parallel$ is a constant $t_\parallel=t$, where t sets the energy scale. However, in leg-2, the hopping amplitude depends on the dimerization parameter $\delta$ and acquires a phase $e^{-i\pi\delta}$ \cite{mellado2025sliding}.

The inter-leg (rung) hopping term between nearest neighbor atoms in the same sublattice is 
\begin{equation}
H_{\perp}
= \sum_{j,\sigma}
t_\perp(j)
\left(
c^\dagger_{j,1,\sigma} c_{j,2,\sigma} + \mathrm{h.c.}
\right),
\end{equation}
The rung tunneling acquires a moiré periodicity inherited from the commensuration between the two legs, a product of the dimerization of leg-2. 

The last term in Eq.\ref{eq:allH} consists of a SOC contribution in each leg that locks spin polarization along the $y$ direction. 

To ensure that the moiré modulation is commensurate, we set $\delta = p/q$ with coprime integers $p$ and $q$, so that both the interleg modulation and the leg-2 dimerization become periodic over $q$ unit cells (for example, for $\delta = \frac{19}{20}$, the corresponding supercell spans $20$ unit cells). The relative mismatch between legs yields a moiré supercell of \(q\) composite cells with \(4q\) sites in total, a moiré reciprocal vector in the reduced Brillouin zone (RBZ) \(b_s=\frac{G_0}{q}=\frac{\pi}{q}\), and a real-space supercell length \(L_s=2\,q\). The intra-leg nearest neighbor hopping, $t$, sets the bandwidth \(W\sim 4t\). Rung tunneling has an average of \(t\). Schrieffer–Wolff perturbation theory \cite{bravyi2011schrieffer} gives rise to a first-harmonic rung tunneling of amplitude \(t_1=(1-\delta)t\) modulated by the envelope $t_\perp(n)=t+t_1\cos\left(\frac{2\pi n}{q}\right)$.  
Structure factors associated with hopping in leg-1 and leg-2 are  respectively \(co(k)=2\cos k\), \(co_\delta(k)=2\cos\!\big(k-\pi\delta\big)\), and the Peierls phase from the dimerization of leg-2 is $\phi(k)=k(1-\delta)+\pi\delta$. In Fourier space, the moiré rung modulation becomes \(t_\perp(k)=1+(1-\delta)\cos\!\big(qk\big)\). Throughout, we choose \(t\) as our energy scale.

In graphene-based heterostructures coupled either to a topological insulator or to a substrate that induces a perpendicular electric field, the dominant proximity-induced SOC is generally of Rashba character \cite{bihlmayer2022rashba}. Microscopically, this Rashba SOC originates from virtual hopping processes, during which an electron experiences a momentum-dependent spin rotation as it moves in the gradient of the interfacial electric field normal to the heterostructure plane \cite{ado2017microscopic,chen2018electric}. In a quasi-one-dimensional lattice description, the Rashba SOC generated by a perpendicular electric field $\mathbf{E}=E_z\hat{z}$ takes the form \cite{bihlmayer2022rashba}
\begin{equation}
H_{\rm SOC}
=
\alpha \sum_j 
\left(
c^\dagger_{j+1}\, i\sigma_y\, c_j
-
c^\dagger_{j}\, i\sigma_y\, c_{j+1}
\right),
\label{eq:soc_realspace}
\end{equation}
where the fermion operators $c_j=(c_{j\uparrow},c_{j\downarrow})^T$, $\uparrow$ and $\downarrow$ indicate the spin-up and spin-down states, respectively, the sigma matrix \(\sigma_y =
\begin{pmatrix}
0 & -i \\
i & 0
\end{pmatrix}
\) acts on spin and $\alpha$ is the SOC strength. This expression embodies how Rashba SOC ties the electron spin to its bonding direction via the interfacial electric field.
Fourier transforming Eq.~\eqref{eq:soc_realspace},
one obtains in leg 1
\begin{equation}
H_{\rm SOC}^{(1)}
=
2i\alpha \sum_k
 \left(\sin k \right)
c_k^\dagger \sigma_y c_k .
\label{eq:soc_kspace}
\end{equation}
In a uniform system such as leg-1, the SOC term in Eq.\ref{eq:soc_kspace} appears as a momentum-odd contribution proportional to $\sin k$, which vanishes at time-reversal invariant momenta and reflects the inversion-symmetry breaking at the interface. 

In leg-2, due to the presence of $\delta$, the spin–orbit coupling term acquires an additional phase factor.
\begin{equation}
H_{\rm SOC}^{(2)}
=
\alpha \sum_j 
\left(
e^{-i\pi\delta}\, c^\dagger_{j+1,2}\, i\sigma_y\, c_{j,2}
-
e^{+i\pi\delta}\, c^\dagger_{j,2}\, i\sigma_y\, c_{j+1,2}
\right),
\label{eq:soc_dimer_real}
\nonumber
\end{equation}
Fourier transforming the previous equation yields
\begin{align}
H_{\rm SOC}^{(2)}
&=
2 i \alpha\sum_k
 e^{i\phi(k)}\sin(k-\pi\delta)\;
c^\dagger_{k,2}\,\sigma_y\, c_{k,2}.
\label{eq:soc_dimer_k}
\end{align}
We define the spinor in the leg/site/spin basis as the following eight-component object,
\(
\Psi_{k}=
(c_{1A\uparrow},c_{1,A\downarrow}, c_{1,B\uparrow},c_{1,B\downarrow},c_{2,A\uparrow},c_{2,A\downarrow},c_{2,B,\uparrow},c_{2,B\downarrow})\).

To model the two surfaces of the heterostructure, we assign opposite helicity to the two legs:
leg~1 carries helicity $+\sigma_y$ (upper surface) and leg~2 carries helicity $-\sigma_y$ (lower surface). The upper two orbitals (leg~1) thus experience a potential $+2i\alpha \sin k\, \sigma_y$, while the lower two orbitals
(leg~2) experience $-2 i \alpha \sin(k-\pi\delta)\,\sigma_y$. The total Hamiltonian, including both the kinetic terms and the opposite-helicity spin–orbit coupling, yields 

\begin{equation}
\label{eq:H}
H(k)=
\begin{pmatrix}
0 & 0 & co(k) & -is(k)& t_\perp(k) &0&0 &0 \\
0 & 0 & is(k) & co(k)&0&t_\perp(k)&0 &0\\
co(k) & -is(k)&0 & 0&0 & 0& t_\perp(k) &0\\
is(k) & co(k)&0 & 0&0 & 0&0&t_\perp(k)\\
t_\perp(k) & 0 & 0 & 0&0 & 0&co_\delta(k)\,e^{i\phi(k)}&is_\delta(k)\,e^{i\phi(k)}\\
0 & t_\perp(k) & 0 & 0 &0& 0&-is_\delta(k)\,e^{-i\phi(k)}& co_\delta(k)\,e^{i\phi(k)}\\
0&0&t_\perp(k)&0&co_\delta(k)\,e^{-i\phi(k)}&is_\delta(k)\,e^{i\phi(k)}&0&0\\
0&0&0& t_\perp(k)&-is_\delta(k)\,e^{-i\phi(k)}&co_\delta(k)\,e^{-i\phi(k)}&0&0\\
\end{pmatrix},
\end{equation}

where $s(k)=2\alpha \sin k$ and $s_\delta(k)=2 \alpha \sin(k-\pi\delta)$.  Eq.\ref{eq:H} is given in the extended Brillouin zone (EBZ), \(\rm k\in (-qG_0,qG_0 )\). For a given $\delta$, the similarity transformation $\Delta=\pi q$  applied to $H(k)$ imposes constraints on p and q that determine the generic global period $2\pi q$ giving rise to EBZ \cite{mellado2025sliding}. 
On the other hand, the reduced first Brillouin zone of the moiré lattice (RBZ) spans the interval \((-b_s/2,b_s/2)\), and the reciprocal superlattice 
\(G_s = \{ m\,b_s\ |\ m\in\mathbb{Z}\}\). 

The two-leg ladder model employed here should be understood as a minimal effective description of the hybridizing degrees of freedom at a graphene/topological-insulator interface. Leg 1 and Leg 2 represent the active electronic states of the graphene $\pi$-orbitals and the TI topological surface states, respectively. While this model simplifies the 2D lattice structure into a quasi-1D geometry to maintain clarity, it preserves the essential physics of the 2D heterostructure.  The relative shift $\delta$ mimics the structural incommensurability or lattice mismatch at the interface, and the momentum odd nature of the Rashba SOC the is naturally captured. The main ingredients omitted are the 2D transverse momentum $k_y$ and the material-specific atomic details of the substrate. However, as demonstrated by the mapping to $\rm Sb_2Te_3$ on graphene ($\delta \approx 19/20$, that corresponds to 5\% of mismatch at the interface, and $\alpha \approx 250$ meV) \cite{munoz2026emergence}, this abstraction is sufficient to predict the robust topological features and helicity fragmentation that would manifest in spectroscopic measurements of real 2D interfaces.
\section{Energy Spectrum and Density of states}
\begin{figure}
  \includegraphics[width=\linewidth]{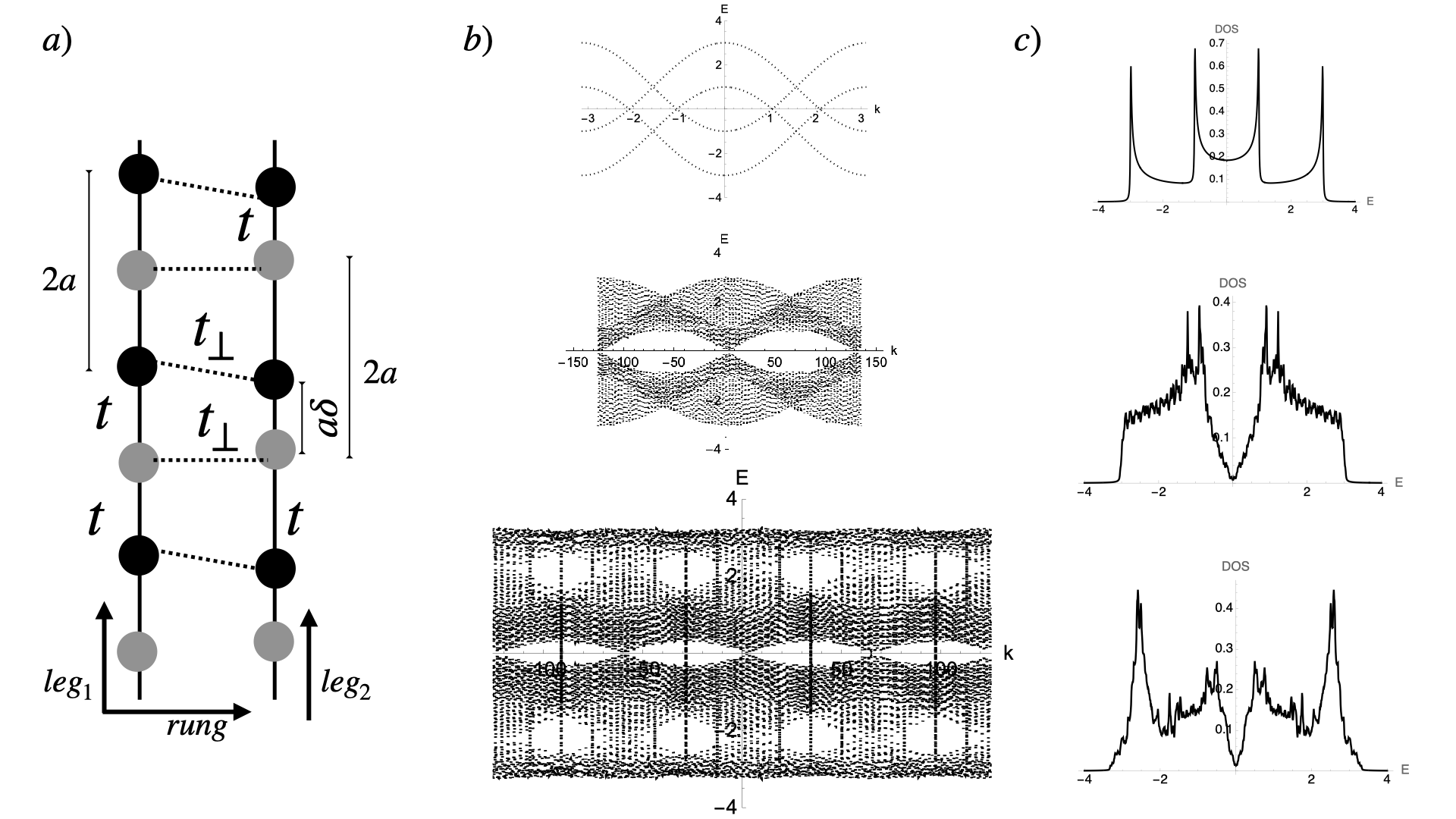}
  \caption{Energy spectrum and model geometry. t sets the energy scale. (a) Schematic of the moiré ladder representing the interface between graphene (leg 1) and a topological insulator (leg 2) (b) Band spectra in the EBZ of $H(k)$, of a spinless system at $\delta=1$ (no mismatch, upper panel) and $\delta=\frac{19}{20}$ (5\% mismatch, middle panel) and of a spinful system with $\alpha=t$ at $\delta=\frac{19}{20}$ (lower panel). $t=\alpha$ represents a strong-coupling regime, with $\alpha\sim 250$ meV at $\rm{Sb}_2Te_3$/graphene interfaces \cite{munoz2026emergence}. The effective periodicity of the spectrum in the EBZ is reduced by a factor of two, reflecting the breaking of pure translational symmetry by momentum-odd Rashba SOC. (c) DOS at T=0 of a spinless system with $\delta=1$ (upper panel), a spinless system with $\delta=\frac{19}{20}$ (middle panel) and a spinful system $\alpha=t$ at $\delta=\frac{19}{20}$ (lower panel).}
  \label{f1}
\end{figure}

The spectrum  of Eq.\ref{eq:H} is exactly solvable \cite{mellado2025excitonic} and is presented in
Figure \ref{f1}.  Fig.\ref{f1}b compares the energy spectrum in the extended Brillouin zone for a modulated system at $\delta=0.95$, in the absence and presence of spin-orbit coupling. In the SOC-free case ($\alpha=0$, middle panel), the spectrum consists of a dense set of folded minibands originating from the moiré modulation \cite{mellado2025excitonic}. The bands remain spin-degenerate throughout the EBZ, and a tiny gap visible around the Fermi energy, E=0, is a direct consequence of \(\delta\), which breaks the translational symmetry.  Minigaps at folded-band crossings are apparent and scale linearly with $t_1$ at small modulation (large $\delta$).  Upon introducing spin–orbit coupling, the overall band structure is preserved, indicating that SOC does not qualitatively alter the moiré folding. However, SOC lifts the spin degeneracy and gaps out a fraction of the band crossings, producing a fine pattern of avoided crossings across the EBZ (lower panel). Due to the moire potential, the spin character is redistributed across the miniband manifold, even though no large global gap opens. In the following analysis, whenever spin–orbit coupling (SOC) is included, we set $\alpha=t$ so as to concentrate on the parameter regime in which proximity-induced SOC constitutes the dominant energy scale. This choice enables a more transparent and systematic characterization of the ensuing helical fragmentation and the corresponding band reconstruction.

In the absence of SOC, the spectrum obeys $E(k)=E(k+G_m)$. Once Rashba SOC is introduced, the Hamiltonian acquires momentum-odd contributions $\propto \sin(k-\pi\delta)\,\sigma_y$, which reverse sign under $k\rightarrow k+G_m$. Consequently, translation by $G_m$ alone ceases to be a symmetry. The Hamiltonian remains invariant only under the combined action of a moiré translation and a $\pi$ spin rotation about the $y$ axis. Because this composite symmetry squares to a translation by $2G_m$, the spectral periodicity in the extended Brillouin zone is effectively halved, accounting for the reduced EBZ period observed in the lower panel of Fig.\ref{f1}(b).
 
We have recently demonstrated that incorporating moiré modulation into a two-leg moiré ladder modifies the associated density of states (DOS) \cite{mellado2025excitonic}. With a perfect match between the legs ($\delta=1$), the DOS shown in the upper panel of Fig.\ref{f1} depicts a pair of van Hove singularities (VHS) at the edges of the bands, a finite density of states at all energies, and another pair of VHS at $ E=\pm t_1$, signaling the coupling between the two legs. Once $\delta\neq 1$, the moiré potential smears out the peaks at the edges of the spectra; the VHS at $E=\pm t$ splits into two peaks about $E\sim\pm t_1$  due to the breaking of inversion symmetry (middle panel); and the DOS $\sim \eta$ at the Fermi energy as the system becomes gapped \cite{ribeiro2015origin}.  Including SOC (lower panel) removes spin degeneracy, opening gaps at many band crossings, redistributing spectral weight, and producing a qualitatively altered DOS. Sharp peaks broaden or split, and new fine structures emerge due to SOC-driven avoided crossings.

To build the Hamiltonian of the system in the RBZ we keep the \(q\) replica momenta \(\{\bar{k}+m G_s\}_{m=0}^{q-1}\) in the Hamiltonian matrix $H_{\mathrm{RBZ}}$. Thus, the block‐diagonal piece becomes
\(
\big[H^{(0)}_{\mathrm{RBZ}}(\bar{k})\big]_{m m'}=
\delta_{m m'}\,H\!\big(\bar{k}+m G_s\big),\) 
\( m,m'=0,\dots,q-1
\). Rung modulation \(\cos(qk)\) connects nearest-neighbor replicas in the RBZ and gives rise to the off-diagonal part \(\big[\delta H_{\mathrm{RBZ}}(\bar{k})\big]_{m,m\pm 1}=\frac{t_1}{2}\,R,\quad
R = \sigma_1 \otimes \mathbb{I}_2\), where $\sigma_j$ is the j-$th$ Pauli matrix. The  \(4q\times4q\) moiré Hamiltonian in the RBZ becomes \cite{mellado2025sliding}
\(H_{\mathrm{RBZ}}(\bar{k})=H^{(0)}_{\mathrm{RBZ}}(\bar{k})+\delta H_{\mathrm{RBZ}}(\bar{k})\).
If \(t_1=0\) the \(4q\) bands split into \(q\) independent copies of the four parent bands evaluated at $\bar{k}+m G_s$. The symmetry of \(H\) is preserved by $ H_{\mathrm{RBZ}}$ \cite{mellado2025sliding}. 

\begin{figure}
  \includegraphics[width=\linewidth]{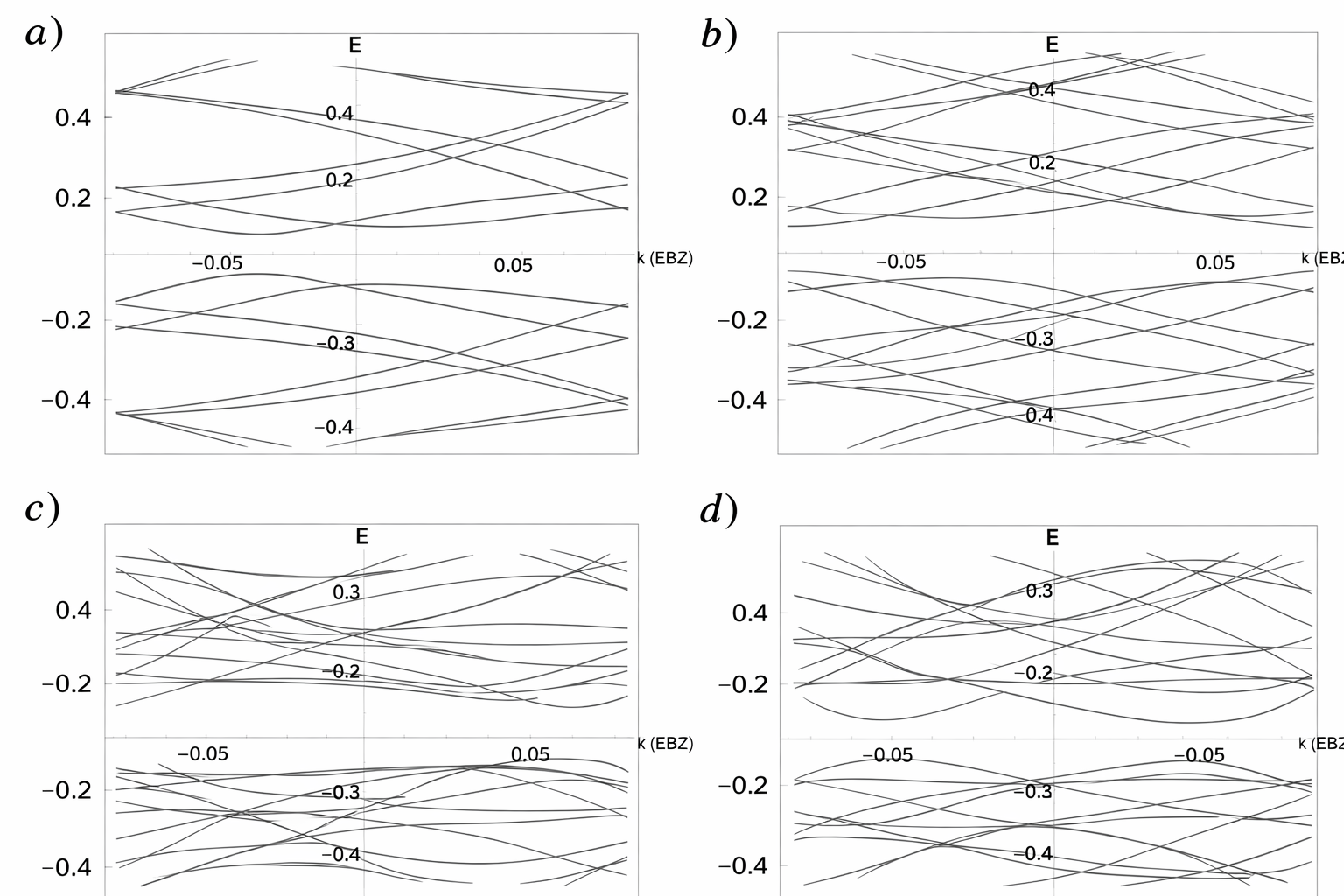}
  \caption{Low-energy miniband structure and Dirac-like crossings. Zoom-in of the moiré miniband spectrum in the RBZ near the Fermi energy ($E=0$). (a,b) Spectrum at $\delta=\frac{19}{20}$ (5\% lattice mismatch) without SOC (a) with $t=\alpha$ (b). (c,d) Spectrum at $\delta=\frac{17}{20}$ (15\% lattice mismatch) without SOC (c) with $t=\alpha$ (d). The inclusion of Rashba SOC leads to the emergence of Dirac-like linear crossings.}
  \label{f2}
\end{figure}

Fig.\ref{f2} shows a zoomed-out view of the energy spectra of the minibands in the reduced Brillouin zone at $\delta=0.95$, that corresponds to 5\% of mismatch between the legs and $\delta=0.85$ (15\% mismatch) in the absence (a,c) and presence (b,d) of spin-orbit coupling. Without SOC, the RBZ spectrum consists of multiple folded minibands that intersect extensively, reflecting both the moiré backfolding and the underlying spin degeneracy. Figs.\ref{f2}(b,d) show that SOC qualitatively restructures the internal organization of the minibands: spin degeneracy is lifted, and a large fraction of the band crossings are replaced by avoided crossings. As a result, the RBZ spectrum becomes more densely interconnected, with miniband branches hybridizing over extended momentum ranges. This reconstruction reflects the entanglement of spin and moiré degrees of freedom induced by SOC and leads to a redistribution of spectral weight across the miniband manifold. 

The Dirac-like crossings observed in Fig.\ref{f2}(d) indicate that the moire modulated SOC not only lifts degeneracies, but also reorganizes the miniband structure into effective two-band sectors with linear dispersion. The case at $\delta=0.85$ illustrates how moiré modulation and SOC jointly induce emergent relativistic behavior in otherwise heavily folded miniband spectra. A symmetry mechanism is expected to safeguard the Dirac-type band crossings in the RBZ spectrum when SOC is included. With SOC present, a pure moiré translation by a reciprocal lattice vector $G_m$ ceases to be a symmetry since the Rashba-like SOC term is odd in momentum and reverses sign under $k \rightarrow k + G_m$. Nevertheless, the Hamiltonian is invariant under the composite operation $\tilde T = T_{G_m}\mathcal R_y(\pi)$, with $\mathcal R_y(\pi)=-i\sigma_y$ denoting a $\pi$ spin rotation about the SOC-defined axis. This combined spin–translation symmetry yields $\tilde T^2 = T_{2G_m}$, thereby halving the fundamental periodicity of the EBZ. $\tilde T$ acts as a quantum number that differentiates miniband eigenstates, such that bands with distinct $\tilde T$ eigenvalues cannot hybridize at the same momentum. As a result, when moiré folding brings bands from different $\tilde T$ sectors into contact, SOC imposes symmetry-protected linear crossings, yielding Dirac-like nodes in the miniband spectrum.

We emphasize that the two distinct types of features exhibited in the spectrum of Fig.\ref{f2} are distinct. The avoided crossings are the primary result of the moiré-modulated tunneling. The structural incommensurability acts as a periodic perturbation that hybridizes states at the boundaries of the reduced Brillouin zone. Where these bands would otherwise intersect, the moiré potential opens a gap, a process that drives the formation of flat minibands. Unlike the avoided crossings, the Dirac like linear crossings remain gapless even at finite SOC. Their robustness is rooted in the composite spin-translation symmetry of the moiré ladder. Because the Rashba interaction is odd in momentum, it couples spin and spatial degrees of freedom such that a translation by a moiré vector followed by a spin flip is a symmetry of the system. At specific $k$-points, where the moiré-induced hybridization would typically open a gap, this symmetry forces the off-diagonal coupling elements to vanish.

\section{Helical modes}
Helical states have been observed in moiré heterostructures featuring proximity induced SOC \cite{khokhriakov2020gate,kiemle2022gate,song2018spin,li2012topological,hoque2024room}. In spin–orbit coupled systems, helicity refers to the projection of the spin degree of freedom onto a momentum and symmetry defined internal axis. For continuum Rashba models, it is typically specified as the eigenvalue of $\hat{\boldsymbol{\sigma}}\cdot(\hat{\boldsymbol{z}}\times\hat{\boldsymbol{k}})$ \cite{manchon2015new}, which quantifies the locking between spin and momentum. In our moiré ladder framework, we define the helicity operator as the difference between the spin densities on the two legs \cite{bychkov1984oscillatory},
\begin{equation}
\Gamma_{\mathrm{hel}}
=
S_y^{(1)} - S_y^{(2)},
\label{eq:helicity_def}
\end{equation}
where
\begin{equation}
S_y^{(\ell)}
=
\sum_k c_{k,\ell}^\dagger \, \sigma_y \, c_{k,\ell},
\end{equation}
with $\sigma_y$ acting in spin space. With this definition, helicity quantifies how unevenly spin polarization along the SOC-determined axis is shared between the two legs. The corresponding operator is antisymmetric under leg exchange, invariant under time reversal, and zero when SOC is absent due to spin-rotation symmetry.

\begin{figure}
  \includegraphics[width=\linewidth]{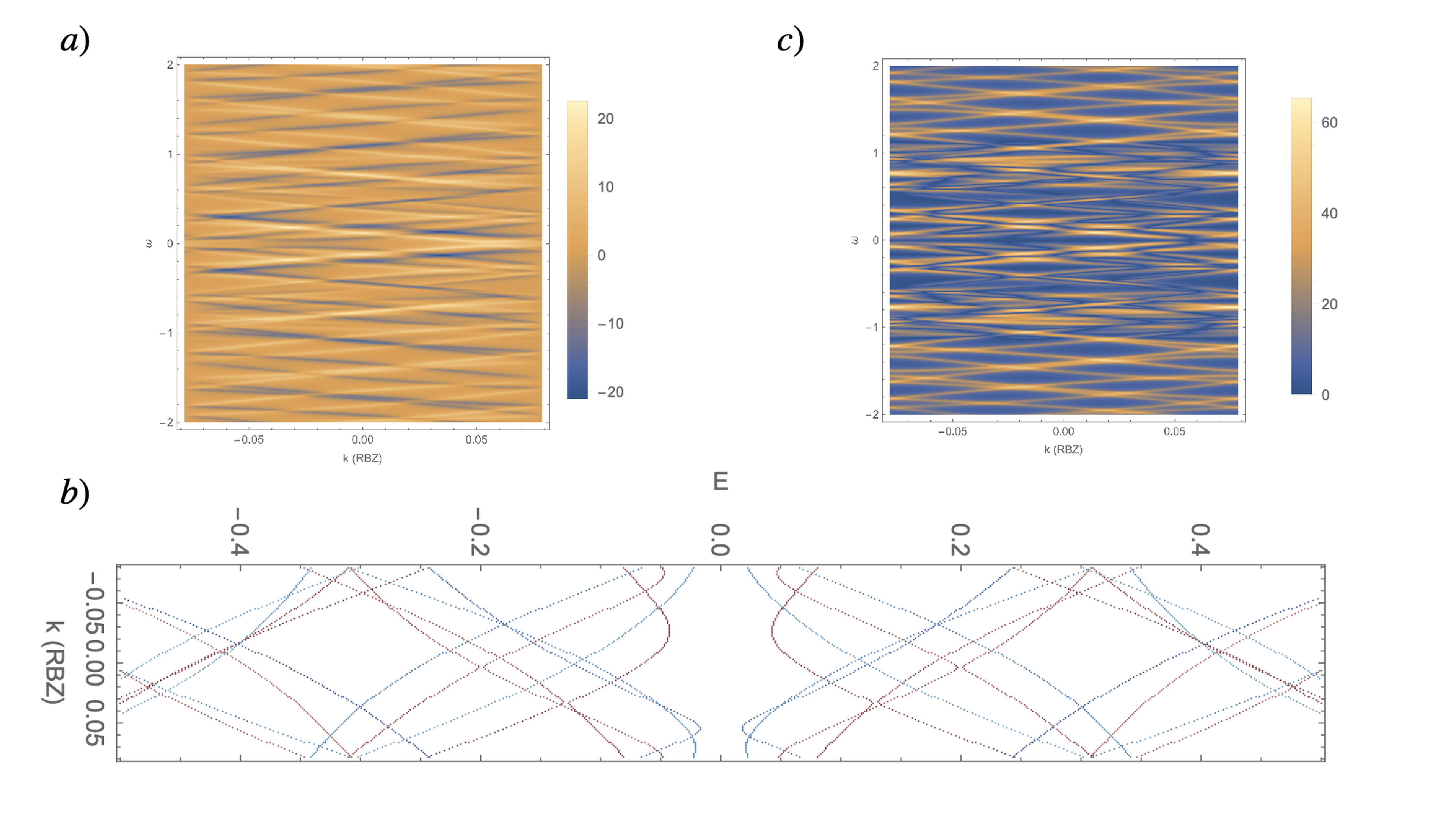}
  \caption{Helicity fragmentation and spectral properties at $5\%$ lattice mismatch, in the strong-coupling regime ($t=\alpha$). (a) Helical spectral function of the system. (b) Detailed view of the moiré miniband spectrum near the Fermi energy ($E=0$). The color mapping reflects the helicity weight of each miniband, illustrating the fragmentation of spin-momentum locking across the moiré manifold. (c) Helicity current spectral function.}
  \label{f3}
\end{figure}

By defining helicity as a difference in leg-resolved spin density, we isolate the specific part of the spin-momentum locking that is modulated by the moiré-induced tunneling. The helicity weight shown in Fig.\ref{f3} represents the degree to which a state is polarized between the two legs in a chiral manner. This is a useful quantity, as it directly relates to the nonlocal spin signals generated when a current is driven through the heterostructure.

In order to find out whether $\Gamma_{\mathrm{hel}}$ appears in single-particle states, we compute the dynamical spectral function in the system using Lorentzian broadening:
\(
A_\Gamma(k,\omega) = \sum_n
|\bra{n,k}{\Gamma}\ket{n,k}|^2
\frac{\eta}{(\omega - E_{n,k})^2 + \eta^2}
\). 
where $\eta$ is a small positive broadening. The helicity-resolved spectral function offers a direct probe of how proximity-induced spin–orbit coupling is redistributed throughout the moiré minibands. Without spin–orbit coupling, spin-rotation symmetry enforces the helicity-resolved spectral function to be zero. Upon introducing SOC, Fig.\ref{f3}a reveals a dense mesh of oblique streaks with alternating blue and white intensities spanning the reduced Brillouin zone. Each streak represents a miniband branch with nonzero helicity weight, where the color encodes both the sign and magnitude of the expectation value of the helicity operator $\Gamma_{\mathrm{hel}}$. The tilted alignment of these features arises from the finite group velocity of the associated states, and the alternating sign directly evidences spin-momentum locking: minibands with opposite dispersion slopes possess opposite helicity, as enforced by time-reversal symmetry.

Unlike uniform Rashba systems, where helicity is restricted to a few well-defined branches, the moiré potential distributes helicity across numerous minibands. The color-coded miniband structure in the RBZ in Fig.\ref{f3}b shows that every miniband branch in the reduced Brillouin zone carries nonzero helicity weight. The simultaneous presence of positive and negative helicity on different branches reflects the odd-momentum character of the SOC and the maintenance of time-reversal symmetry. While the case $\alpha = t$ is chosen here to highlight the band reconstruction in the strong-coupling limit, the features remain qualitatively robust for realistic graphene/$\text{Sb}_2\text{Te}_3$ interfaces where $\alpha<t$ and the proximity-induced SOC $\alpha$ is typically on the order of $\sim 250$ meV \cite{munoz2026emergence}. The redistribution of helicity weight shown in (b) serves as a unique fingerprint for spin-resolved ARPES and nonlocal transport measurements. 

Examining the helicity current is particularly relevant in systems with proximity-induced spin–orbit coupling, where spin–momentum locking enables these internal spin textures to influence quasiparticle dynamics even when there is no net charge or spin flow \cite{sinova2004universal,rashba2003spin}. The helicity current operator is defined as the symmetrized product of the velocity operator and the helicity density,
\begin{equation}
J_{\mathrm{hel}}(k)
=
\frac{1}{2}\left\{ v(k), \Gamma_{\mathrm{hel}} \right\},
\label{eq:helicity_current_def}
\end{equation}
where $v(k)=\partial H(k)/\partial k$ denotes the band velocity operator. Fig.\ref{f3}c shows a dense array of tilted streaks with alternating signs and magnitudes. Each streak represents a dispersive miniband that carries a finite helicity current, whose sign is fixed by the interplay between the group velocity and helicity polarization. Time-reversal symmetry requires the presence of counterpropagating channels with opposite helicity current, producing the characteristic interlaced pattern throughout the reduced Brillouin zone. Analogous to helicity density, moiré-driven miniband hybridization disperses the helicity current over the full spectrum instead of confining it to a few Rashba-like branches.

Fig.\ref{f3}b shows that no minibands remain helicity-silent. Therefore, fluctuations in the helicity order parameter can be derived from an extensive set of particle-hole excitations involving all minibands, making the system intrinsically prone to helical collective behavior.
\section{Helical fluctuations}
\begin{figure}
  \includegraphics[width=\linewidth]{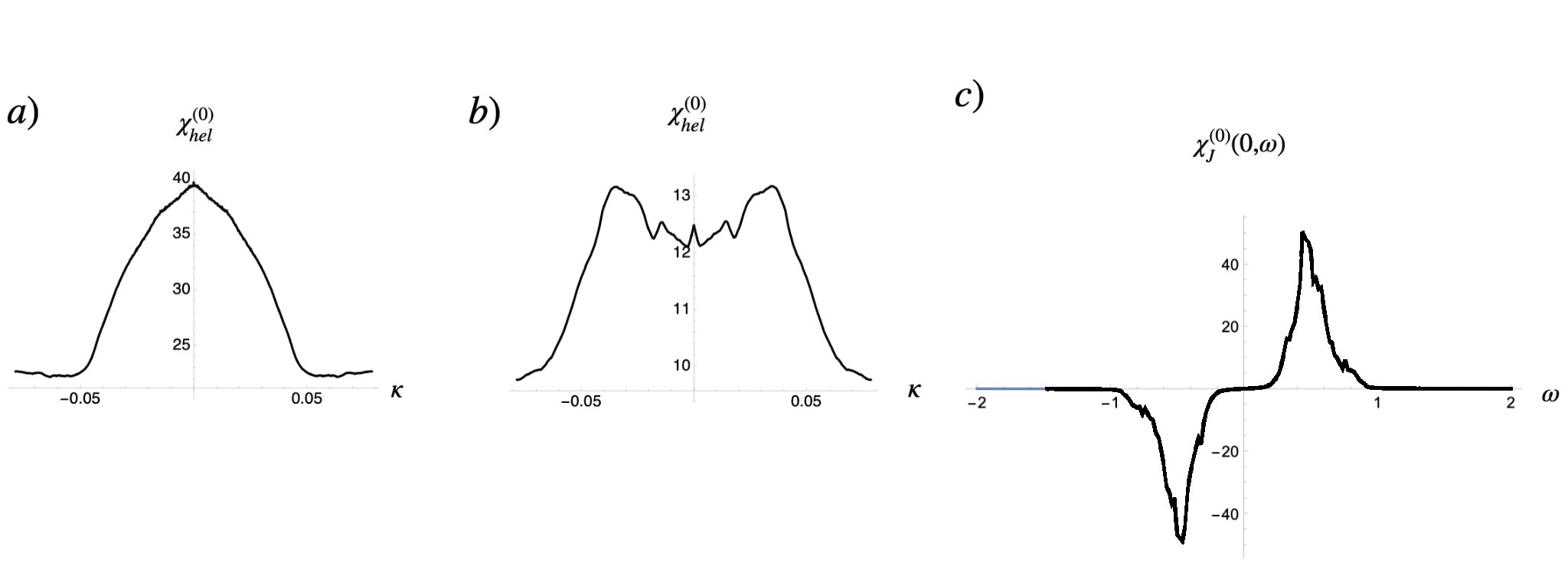}
  \caption{Helicity susceptibility and collective fluctuations. (a,b) Lindhard susceptibility of the moiré system in the strong-coupling regime ($t=\alpha$) and $T=0$ for (a) $5\%$ lattice mismatch and (b) $15\%$ lattice mismatch. The sharp peaks signify enhanced helical fluctuations at the bare-response level, suggesting a tendency toward collective helical states. (c) Dynamical susceptibility of the helicity current operator at $t=\alpha$, $T\sim 250$ K, and $15\%$ lattice mismatch.}
  \label{f4}
\end{figure}

The tendency toward instability in the helicity channel is captured by the associated two-particle response function. To characterize how strongly the system favors forming a helicity modulation with wavevector $\kappa$, we apply an external (generalized) field $h(\kappa)$ that couples linearly to the helicity density operator $\Gamma_{\mathrm{hel}}(\kappa)$ \cite{fetter2012quantum}. The resulting Hamiltonian takes the form
\begin{equation}
H_h \;=\; H \;-\; \sum_\kappa h(\kappa)\,\Gamma_{\mathrm{hel}}(-\kappa),
\label{eq:Hfield}
\end{equation}
where $H$ denotes the unperturbed single-particle Hamiltonian, and $\Gamma_{\mathrm{hel}}(\kappa)$ represents the Fourier component of the helicity density. Within linear response theory, the induced expectation value is connected to the external field via
\begin{equation}
\delta\langle \Gamma_{\mathrm{hel}}(\kappa)\rangle
\;=\;
\chi_{\mathrm{hel}}(\kappa,\omega)\,h(\kappa,\omega),
\label{eq:linresp_def}
\end{equation}
with the susceptibility defined by the Kubo formula \cite{mahan2000strong}
\begin{equation}
\chi_{\mathrm{hel}}(\kappa,\omega)
\;\equiv\;
-i\int_0^{\infty} dt\, e^{i(\omega+i0^+)t}\,
\Big\langle\big[\Gamma_{\mathrm{hel}}(\kappa,t),\Gamma_{\mathrm{hel}}(-\kappa,0)\big]\Big\rangle .
\label{eq:kubo_retarded}
\end{equation}
Here $\Gamma_{\mathrm{hel}}(\kappa,t)=e^{iH_0 t}\Gamma_{\mathrm{hel}}(\kappa)e^{-iH_0 t}$ is the Heisenberg operator, and $\langle\cdots\rangle$ denotes a ground-state expectation value.
The static helicity susceptibility corresponds to the zero-frequency limit,
\begin{equation}
\chi_{\mathrm{hel}}(\kappa)
\;\equiv\;
\chi_{\mathrm{hel}}(\kappa,\omega\!=\!0)
\;=\;
\lim_{\omega\to 0}\chi_{\mathrm{hel}}(\kappa,\omega).
\label{eq:chi_static_def}
\end{equation}
With the sign convention used in this work (so that a positive $\chi$ corresponds to an enhancement of helicity fluctuations), we finally write \cite{mahan2000strong}

\begin{equation}
\chi_{\mathrm{hel}}^{(0)}(\kappa)
=
-\frac{1}{N_k}
\sum_{k,n,m}
\frac{f\!\left(E_{n}(k)\right)-f\!\left(E_{m}(k+\kappa)\right)}
{E_{n}(k)-E_{m}(k+\kappa)}
\left|
\langle n,k \vert \Gamma_{\mathrm{hel}} \vert m,k+\kappa \rangle
\right|^2 ,
\label{eq:lindhard_analytic}
\end{equation}

where $E_{n}(k)$ are the miniband energies, and $f(E)$ is the Fermi-Dirac distribution at finite temperature $T$. A pronounced maximum in $\chi_{\mathrm{hel}}^{(0)}(\kappa)$ indicates collective fluctuations in the helicity channel and reveals the characteristic momentum profile of the leading helical mode. This determines whether the noninteracting band structure shaped by moiré modulation and proximity-induced SOC is inherently inclined toward helical instabilities, thereby supplying the microscopic basis for helical phases to develop once additional symmetry-breaking fields or interactions are applied.
\\
Fig.\ref{f4}(a,b) presents the helical susceptibility for $\delta=\frac{19}{20}$ and $\delta=\frac{17}{20}$ at $\alpha=t$. For $\delta=0.95$, a pronounced maximum appears at $\kappa=0$, indicating a propensity toward a uniform helical phase. Lowering the modulation to $\delta=0.85$ substantially alters the profile of the static helicity susceptibility $\chi_{\mathrm{hel}}(\kappa)$. For stronger modulation, the uniform mode at $\kappa=0$ persists, but the susceptibility exhibits symmetric peaks at $\pm \kappa^*$, demonstrating that the leading helical fluctuations are shifted to a finite wave vector. This behavior marks a tendency for a transition from an essentially uniform helicity response to an emerging helical density-wave state.

In order to probe frequency resolved fluctuations of helicity transport \cite{burkov2010spin,sinova2004universal}, we consider the dynamical susceptibility associated with the helicity current operator.

\begin{equation}
\chi^{\mathrm{hel}}_{J}(\kappa,\omega)
=
\sum_{k,n,m}
\frac{f(E_{n}(k))-f(E_{m}(k+\kappa))}
{\omega + i\eta + E_{k n}-E_{k+\kappa,m}}
\,
\left|
\Gamma^{\mathrm{hel}}_{J,nm}(k,k+\kappa)
\right|^{2},
\label{eq:chiHelJ}
\nonumber
\end{equation}

which is the retarded current–current correlator.  The vertex
$\Gamma^{\mathrm{hel}}_{J,nm}(k,k+\kappa)
=
\langle n,k|\hat J_{\mathrm{hel}}(k)|m,k+\kappa\rangle$
is the momentum dependent helicity current matrix element. The imaginary part of the susceptibility,
$\Im \chi^{\mathrm{hel}}_{J}(\kappa,\omega)$, represents the absorption spectrum of helicity current fluctuations \cite{fetter2012quantum} and quantifies the phase space of particle–hole excitations \cite{burkov2010spin}, thereby capturing both the band-structure kinematics and the internal helicity texture of the minibands.

Figure~\ref{f4}(c) shows $\Im \chi^{\mathrm{hel}}_{J}(q=0,\omega)$ computed at
$\delta=0.85$ for $\alpha=t$, and $T=0.1$. For a realistic graphene/$\text{Sb}_2\text{Te}_3$ interface, the temperature $T=0.1t$ corresponds to $\sim 250$ K, indicating that these helical signatures are robust well above cryogenic temperatures. The mirror feature at negative frequencies follows from the antisymmetry of the imaginary part. The peaks at low positive frequencies arise from low-energy interband particle–hole excitations between moiré minibands, whose energies are brought close by the modulation and further reshaped by spin–orbit coupling. The helicity current vertex amplifies these transitions, producing a finite frequency response even at zero momentum. The peaks observed in the Lindhard and dynamical susceptibilities (Fig.\ref{f4}) indicate a significant enhancement of the response in the helicity channel. At the bare-response level presented here, these features should be interpreted as enhanced helical fluctuations rather than a definitive collective phase.

The moiré-induced band reconstruction provides a microscopic mechanism that increases the density of states at specific nesting vectors, creating a strong propensity toward collective helical order.While a full treatment of electron–electron interactions is ultimately necessary to unambiguously characterize the microscopic structure of the broken-symmetry phase, the present analysis shows that the moiré-modulated Rashba spin–orbit coupling provides the requisite fluctuation channels to facilitate helical fluctuations at the interface. As noted in our related study on moiré ladders \cite{mellado2025sliding}, similar susceptibility enhancements in quasi-1D systems are known to precede the formation of charge-density-wave and sliding-phason phases.
\section{Conclusions and Outlook}
To explore how proximity-induced spin–orbit coupling in TI/graphene interfaces relates to the moiré potentials arising at heterostructure interfaces, we have studied a toy model of a moire ladder system, in which the spin–orbit interaction acquires the moire modulation through a slight dimerization of one ladder leg.

Without SOC, the spectrum is formed by spin-degenerate moiré minibands whose periodicity is fixed by the superlattice folding. Incorporating SOC removes the spin degeneracy and reduces the effective spectral periodicity in the EBZ by a factor of two. This reduction originates from the momentum-odd Rashba spin–orbit coupling, which breaks the equivalence of states related by the original moiré translation symmetry, and constitutes a direct manifestation of the moiré potential entangling the spatial periodicity with the spin–momentum locking at the interface. In this way, the system effectively realizes a new synthetic periodicity for the helical states.

The moiré parameter $\delta$ compresses the electronic spectrum into narrow minibands and increases the density of states near the Fermi level. Moiré-modulated SOC entangles spin, sublattice, and leg degrees of freedom, reshaping the miniband structure in momentum space instead of inducing a simple rigid spin splitting. We found that helicity weight spreads over a dense manifold of moiré minibands, forming an extended network of helicity-carrying states rather than being confined to a few Rashba-like branches. Because positive and negative helicity sectors coexist at the same energy, the total helicity current cancels in equilibrium, despite a large local helicity-resolved spectral weight. The helicity spectral function, therefore, captures internal spin-leg textures and fluctuations rather than transport. 

The appearance of Dirac-like miniband crossings at finite SOC shows that moiré heterostructures can host relativistic quasiparticles via band reconstruction, even when the underlying bands are nonrelativistic. Consequently, helicity-resolved static susceptibilities exhibit sharp peaks, revealing strong collective helicity fluctuations already at the bare-response level.

The robustness of the moiré-folded helical states is a central feature of our model. While we have focused on the specific case $t = \alpha$ to clearly visualize the spectral reconstruction, the underlying physical phenomena are qualitatively generic. The reduction of spectral periodicity by a factor of two is a symmetry-protected consequence of the momentum-odd Rashba interaction. As such, it persists for any non-zero $\alpha$. While the size of the avoided crossings and the helicity weight distribution scale linearly with $\alpha$ in the weak-coupling regime, the fragmentation of the spin-texture remains a robust hallmark of the moiré-modulated interface. While a larger supercell leads to a denser set of minibands and more frequent avoided crossings, the mechanism of band reconstruction remains unchanged. The Dirac-like crossings, in particular, are protected by composite spin-translation symmetry and do not require fine-tuning of the structural mismatch. As shown in the susceptibility analysis, the helical fluctuations remain well-defined at $T=0.1t$. This indicates that the predicted signatures are not restricted to the ground state but could be accessible within the thermal broadening typical of room-temperature experimental conditions.

From a materials science perspective, the present model offers a pertinent description of graphene interfaced with a three-dimensional topological insulator, such as $\text{Sb}_2\text{Te}_3$. In such van der Waals heterostructures, interfacial orbital hybridization generates a substantial Rashba-type spin-orbit coupling, while the intrinsic lattice mismatch of magnitude $\delta \approx 5\%$ gives rise to a long-wavelength moiré modulation of this coupling. 

Regarding experimental characterization, the resulting spin-split moiré minibands and the associated avoided band crossings should be directly accessible via spin- and angle-resolved photoemission spectroscopy. Moreover, the fragmentation of spin helicity across different minibands is, in principle, observable using spin-resolved spectroscopic techniques. The pronounced helicity fluctuations already evident at the bare-response level indicate that spin–charge interconversion, as probed in nonlocal transport measurements, will exhibit a strong sensitivity to the precise position of the Fermi level within the moiré miniband manifold. These phenomena emerge generically from the concurrent presence of spin-orbit coupling and moiré superlattice modulation and do not depend on fine-tuning of microscopic parameters.

Our findings show that moiré heterostructures constitute an advantageous setting where proximity-induced SOC is not simply transferred from the interface, but significantly enhanced and reshaped through miniband hybridization, thereby strongly favoring the emergence of helical states when additional symmetry-breaking fields or interactions are applied.

% Acknowledgements
\medskip
\textbf{Acknowledgements} \par %delete if not applicable))
The author acknowledges support from the Fondo Nacional de Desarrollo Científico y Tecnológico (Fondecyt) under Grant No. 1250122.

% References
\medskip

% Use the following code if you wish to generate your bibliography with BibTeX;
% replace the string "MSP-template" below with the name(s) of
% the BibTeX data base(s) you want to use.
% The resulting bibliography-output (the content of the .bbl file)
% must be pasted back into this file before submission.
% Please also include your BibTeX data base file(s) in your submission
% so that we can re-run BibTeX if necessary.
%
%\bibliographystyle{MSP}
%\bibliography{helical}

% Table of contents entry should be 50 - 60 words long
% Image should be 55 mm broad and 50 mm high or 110 mm broad and 20 mm high

\end{document}